\documentclass[12pt]{article}
\usepackage{amssymb,latexsym,epsfig}

\begin{document}

\newcommand{\eref}[1]{(\ref{#1})}
\newcommand{\sref}[1]{\S\ref{#1}}
\newcommand{\tref}[1]{Table~\ref{#1}}
\newcommand{\fref}[1]{Figure~\ref{#1}}
\newcommand{\cref}[1]{Chapter~\ref{#1}}
\newcommand{\bcenter}{\begin{center}}
\newcommand{\ecenter}{\end{center}}
\newcommand{\beq}{\begin{equation}}
\newcommand{\eeq}{\end{equation}}
\newcommand{\be}{\begin{equation}}
\newcommand{\ee}{\end{equation}}
\newcommand{\bea}{\begin{eqnarray}}
\newcommand{\eea}{\end{eqnarray}}
\newcommand{\bean}{\begin{eqnarray*}}
\newcommand{\eean}{\end{eqnarray*}}
\newcommand{\ba}{\begin{array}}
\newcommand{\ea}{\end{array}}
\newcommand{\ben}{\begin{enumerate}}
\newcommand{\een}{\end{enumerate}}
\newcommand{\bi}{\begin{itemize}}
\newcommand{\ei}{\end{itemize}}
\newcommand{\bd}{\begin{description}}
\newcommand{\ed}{\end{description}}
\def\IC{\mathbb{C}}
\def\II{\mathbb{I}}
\def\IQ{\mathbb{Q}}
\def\IR{\mathbb{R}}
\def\IF{\mathbb{F}}
\def\IZ{\mathbb{Z}}
\def\IO{\mathbb{O}}
\def\IP{\mathbb{P}}
\def\IH{\mathbb{H}}
\def\IK{\mathbb{K}}

\newcommand{\pa}{\partial}
\newcommand{\vev}[1]{\langle #1 \rangle}
\newcommand{\ket}[1]{|#1\rangle}
\newcommand{\bra}[1]{\langle #1 |}
\newcommand{\ack}[1]{[{\bf Ack!: #1}]}
\def\tr{{\rm tr}}
\def\Tr{{\rm Tr}}
\newcommand{\diff}[2]{\frac{\partial #1}{\partial #2}}
\def\nn{\nonumber}
\newcommand{\Real}[1]{\mathfrak{Re}(#1)}
\newcommand{\Imag}[1]{\mathfrak{Im}(#1)}
\def\ie{{\em i.e.},\ }
\def\eg{{\em e.g.},\ }
\def\viz{{\em viz},\ }
\def\cf{{\em cf.}\ }

\def\jfn{$j$-function}
\def\BC{\mathbb{C}}
\def\BR{\mathbb{R}}
\def\BZ{\mathbb{Z}}
\def\CD{{\cal D}}
\def\CO{{\cal O}}
\def\CM{{\cal M}}
\def\CN{{\cal N}}
\def\CW{{\cal W}}

\newcommand{\BP}{\mathbb{P}}
\newcommand{\CS}{{\cal S}}
\newcommand{\Wtree}{W_{\rm tree}}
\newcommand{\Weff}{W_{\rm eff}}
\newcommand{\bPhi}{\overline\Phi}
\newcommand{\bD}{\overline D}
\newcommand{\Vol}{{\rm Vol}}

\newcommand{\Mon}{\mathfrak{M}}

\def\w3{e^{\frac{\pi i}{3}}}
\newcommand{\vpn}{Voiculescu polynomials}
\newcommand{\id}{{1\!\!\!1}}
\newcommand{\vv}[1]{\bra{0}{#1}\ket{0}}


\thispagestyle{empty}
{\flushright{\small{UPR-1048-T\\VPI-IPPAP-03-12\\{\tt 
hep-th/0307293 }\\}}}

\vspace{.3in}
\begin{center}
\bf\LARGE{Modular Matrix Models}
\end{center}
 
\vspace{.2in}
\begin{center}
{\large Yang-Hui He$^\dagger$ and Vishnu Jejjala$^*$\\}
\normalsize{yanghe@physics.upenn.edu, vishnu@vt.edu \\}

\vspace{.2in}
{\it 
$^\dagger$Department of Physics,\\
David Rittenhouse Laboratories,\\
The University of Pennsylvania,\\
209 S.\ 33$^{rd}$ St.,
Philadelphia, PA 19104-6396, U.S.A.\\

\vspace{.1in}
$^*$Institute for Particle Physics and Astrophysics,\\
Physics Department, Robeson Hall, \\ 
Virginia Tech,\\
Blacksburg, VA 24061-0435, U.S.A.\\
}

\vspace{.2in} 
\end{center}

\vspace{0.1in}
\begin{abstract}

Inspired by a formal resemblance of certain $q$-expansions of modular forms
and the master field formalism of matrix models
in terms of Cuntz operators, we construct
a Hermitian one-matrix model, which we dub the ``modular matrix model.''
Together with an $\CN=1$ gauge theory and a special Calabi-Yau geometry,
we find a modular
matrix model that naturally encodes the
Klein elliptic $j$-invariant, and hence,
by Moonshine, the irreducible representations of the Fischer-Griess
Monster group.

\end{abstract}

\newpage
~\\ ~\\ ~\\
\tableofcontents
\newpage

\begin{flushright}
{\it
{\Large W}enn nur ein Traum das Leben ist,\\
Warum denn M\"uh' und Plag'?\\
Ich trinke, bis ich nicht mehr kann,\\
Den ganzen, lieben Tag!}
\end{flushright}

\vspace{0.5in}

\section{Introduction} \label{sec:intro}

The recent resurrection of the old matrix model in string theory has
invited the community to re-examine many intricate properties of random
matrix integrals under a new light.
The technology of using fat matrix Feynman diagrams to cover the Riemann
surfaces associated to the string worldsheet
is once more pertinent.
Non-perturbative information about a wide class of $\CN =1$ gauge theories
is ascertained from partition functions of bosonic matrix
models \cite{DV1,DV2,DV3}.
Moreover, the elegant relation between the spectral curve associated to
the eigenvalues of a random matrix and the special Calabi-Yau geometry that
is used to engineer the supersymmetric gauge theory has been pointed out.

A somewhat parallel phenomenon occurred in mathematics some two decades ago.
McKay and Thompson observed \cite{th} that the coefficients in the
Fourier expansion of Klein's modular invariant function $j(q)$ are simple
linear combinations of the dimensions of the irreducible representations
of the then freshly-conjectured Monster sporadic group.
This observation prompted the ``Monstrous Moonshine'' conjectures \cite{CN}
that compelled mathematicians to re-investigate the century-old \jfn\ with a
modern eye.
The final {\em tour de force} proof of the Moonshine conjectures by
Borcherds \cite{bo} would employ methods that Frenkel, Lepowsky, and
Meurman \cite{FLM} borrowed from vertex algebra techniques arising from
two-dimensional conformal field theory.

\newcommand{\const}{\rm const.}
\newcommand{\Leech}{\rm Leech}
\newcommand{\LLeech}{{\Lambda_{\Leech}}}

The \jfn\ is not new to string theory.
The modular invariant torus partition function for bosonic closed string
theory on the 24-dimensional torus obtained by quotienting $\IR^{24}$ on
the Leech lattice\footnote{The Leech lattice $\LLeech$ is the unique even,
self-dual lattice in $24$ dimensions with no points of length-squared two.
See Ref.\ \cite{CS} for details.} is exactly determined by the \jfn\
\cite{DGH}:
\be
Z_{\Leech}(q) = \frac{\Theta_\LLeech(q)}{\eta(q)^{24}}
= \frac{\sum\limits_{\beta\in\LLeech} q^{\frac{1}{2}\beta^2}}{\eta(q)^{24}} 
= j(q) + \const,
\ee
where $\eta(q) = q^{1/24} \prod\limits_{n=1}^\infty (1 - q^n)$ is the
Dedekind $\eta$-function.
The central charge of the conformal field theory (CFT) fixes the constant
term in $Z(q)$.
The Monster sporadic group is itself the automorphism group of (the $\IZ_2$
orbifold of) this CFT.

We suggest that the parallels between recent trends in string theory
and Moonshine run deeper still.
Inspired by McKay's often daring and insightful speculations, in this
paper we formulate an observation that relates the matrix model
description of field theories to the Klein invariant \jfn.
We believe that this interrelation is non-trivial and itself hints at
intricate and profound connexions among modular forms, simple groups,
$\CN=1$ gauge theories, and Calabi-Yau geometry, as well as to random
matrix theory.
String theory is a unifying principle that blends these disparate notions
together.

The observation begins with the form of the $q$-expansion of the
\jfn, which is the following:
\beq
j(q) = \frac{1}{q} + b_0 + b_1 q + b_2 q^2 + \ldots.
\eeq
This is reminiscent of an object in matrix models.
There is a famous formulation of matrix models known as the ``Master Field''
\cite{Ed}, in which all correlation functions in the model can be
computed without recourse to complicated integrals over infinite dimensional
random matrices.
For the (Hermitian) one-matrix model, correlators are encoded in the
master field $\hat{M}$, which is an operator acting on a certain Fock
space of free probability.
This master field is expandable in a basis of so-called Cuntz
operators $\{a, a^\dagger\}$ as
\beq
\hat{M} = a + m_0 + m_1 a^\dagger + m_2 (a^\dagger)^2 + \ldots.
\eeq

And so an immediate task is evident.
What is the matrix model whose master field is the Klein invariant \jfn?
What is its potential, its free-energy, and its density of eigenvalues?
Utilizing further the Dijkgraaf-Vafa dictionary, what then are the $\CN=1$
gauge theory whose Seiberg-Witten curve and the geometrically engineerable
Calabi-Yau threefold whose special geometry encode this modular
invariant?
All these various players on the stage would naturally encode the
$q$-coefficients of $j(q)$, and hence by Moonshine, the irreducible
representations of the Monster.

Of course, the formal similarity we have noted applies if we take the
Taylor series for a generic function $f(q)$ and compare $f(q)/q$ to the
Cuntz-expansion for the Master field. However, the resolvent thus
defined may not have the correct branch-cut structure to grant
us a consistent matrix model, even if it could be analytically
determined. Happily, the \jfn\ does satisfy these requisites.
We have chosen to concentrate on $j(q)$ 
because it
is the primitive modular function in that any 
other meromorphic function that is modular invariant can be constructed
from it.
To consider how modular invariant functions translate to matrix models,
it therefore makes sense to treat $j(q)$ as the exemplar.
Furthermore, the intimate relation between the \jfn\ and the Monster group
may suggest a new arena for examining the properties of the largest of 
sporadic groups using string theoretic and random matrix techniques. 
A realization of the Monster in terms of geometry and
physics is by itself interesting.

The organization of this article is as follows.
Section \ref{sec:backg} will briefly review the pieces of our puzzle,
\viz the elliptic \jfn, the master field formulation of the Hermitian
one-matrix model, and the Dijkgraaf-Vafa correspondence.
Next, in Section \ref{sec:mmm}, we construct explicitly the matrix model
whose master field is $j(q)$.
We shall dub this the {\em modular matrix model}.
The potential for the model will enjoy the property that 
\beq 
V'(z) \sim j^{-1}(q) + \frac{1}{j^{-1}(q)}, \quad q = e^{2 \pi i z},
\eeq
for $ z \in [1,\infty) \subset \IR$.
We compute the relevant observables in this matrix model.
Then, we construct the $\CN=1$ gauge theory, its full non-perturbative
superpotential, and the Calabi-Yau on which the theory may be geometrically
engineered.
In due course, we shall find a hyperelliptic curve which encodes the
$q$-coefficients.
This is perhaps related to a conjecture of Lian and Yau regarding the
mirror map of a class of K3 surfaces as $1/j(q)$.
We conclude with a discussion of various prospects that await our
investigation. 

\section{Dramatis Person\ae} \label{sec:backg}

In this section, we introduce three characters who shall play important
r\^{o}les in our narrative, \viz the Klein $j$-invariant, the
master field formalism of Hermitian one-matrix models, and the Dijkgraaf-Vafa
correspondence between supersymmetric gauge theories and matrix models.
As all these notions figure prominently elsewhere in the literature,
this section is included as a concise review.

%
\subsection{The Klein Invariant \jfn} \label{sec:jfn}
We begin by introducing the most important function invariant under
the modular group:
the Klein invariant function $j(q)$, with $q := e^{2\pi i z}$.
The \jfn\footnote{Though we are following the literature
in calling $j(q)$ the {\em Klein} invariant \jfn, the attribution is in fact 
less straightforward.
The earliest reference to the modular \jfn\ of which we are aware is C.\
Hermite, ``Sur la r\'esolution de l'\'equation du cinqui\`eme degr\'e,''
{\em Comptes Rendus}, {\bf 46} (1858).
Dedekind and Kronecker anticipated Klein as well.}\ 
is important because it is the {\em unique} modular function in that all
meromorphic functions invariant under $SL(2,\BZ)$ are meromorphic functions
of $j(q)$. In terms of the
Jacobi $\vartheta$-functions, the \jfn\ is defined as
\bea
j(q) & := & 1728\ J(\sqrt{q}), \label{j} \\
J(q) & := & \frac{4}{27}\,\frac{(1-\lambda(q)+\lambda(q)^2)^3}
{\lambda(q)^2(1-\lambda(q))^2}, \label{J} \\
\lambda(q) & := & \left( \frac{\vartheta_2(q)}{\vartheta_3(q)} \right)^4.
\eea
The function $J(q)$ is referred to as Klein's absolute invariant.

The \jfn\ is a modular invariant in the sense that
under the $SL(2,\IZ)$ transformation
\beq
z \mapsto z' = \frac{a z + b}{c z + d}, 
~~~~~ ad-bc = 1,
~~~~~ a,b,c,d \in\IZ,
\eeq
$j(q = e^{2\pi i z}) = j(q' = e^{2\pi i z'})$.
Thus, to study the behavior of the \jfn, it suffices to look at the
fundamental domain defined by ${\cal H} / SL(2; \IZ)$, where
${\cal H}$ is the upper half $z$-plane, $\{\Imag z > 0 \}$.
The \jfn\ maps this domain to the complex numbers:
\beq
j(e^{2\pi i z}) : {\cal H}/ SL(2; \IZ) \rightarrow \IC.
\eeq

We briefly relate a number of other properties regarding the \jfn\ 
\cite{mathworld}:
\begin{itemize}
\item The \jfn\ is meromorphic on ${\cal H} / SL(2; \IZ)$.
\item $j(q)$ is an algebraic number, which is rational or integer at
special values of $q$.
\item The $q$-expansion of the modular invariant \jfn\ is
\bea \label{fourier}
j(q) & = & q^{-1} + 744 + 196884\, q + 21493760\, q^2 + 864299970\, q^3 + \\
&& 20245856256\, q^4 + 333202640600\, q^5 + 4252023300096\, q^6 +
\ldots, \nn 
\eea
which is convergent on the fundamental domain.
Indeed, on the upper-half plane, $\Imag{z}>0$, and so
$q = e^{2\pi i\Real{z}} e^{-2\pi \Imag{z}}$ has an exponential decay
which overwhelms the growth of these integer $q$-coefficients.
\item The coefficients in the $q$-expansion \eref{fourier} are
remarkable. They
are simple linear combinations of the dimensions of the 
irreducible representations of the
Fischer-Griess Monster sporadic group (also, the ``Friendly Giant''):
\[
\ba{rcl}
\mbox{\jfn} & \qquad  & \mbox{Monster} \nn \\
\hline
196884 &=& 1 + 196883, \nn \\
21493760 &=& 1 + 196883 + 21296876, \nn \\
864299970 &=& 2\cdot 1 + 2\cdot 196883 + 21296876 + 842609326, \nn \\
\ldots
\ea
\]
This beautiful observation due to McKay and Thompson \cite{th},
subsequently dubbed by Conway and Norton \cite{CN} as
``Monstrous Moonshine,'' was proved by Borcherds \cite{bo}.
It is truly arresting that the most important modular function is related
to the largest simple sporadic group.
(See Ref.\ \cite{tg} for a fascinating tangled history of the Monster
sporadic group and the modular \jfn.)
\end{itemize}

\subsection{The One-Matrix Model} \label{sec:1mm}
Our next ingredient comes from the physics of $(0+1)$-dimensional matrix
quantum mechanics.
We briefly recall some relevant facts about the zero-dimensional, one-matrix
model.
Our discussion follows the conventions of the review Ref.\ \cite{FGZ}.
Let $\Phi$ be an $N\times N$ Hermitian matrix,\footnote{Such a Hermitian
matrix has $N$ {\em real} eigenvalues.
One can equally well think of matrices $\Phi$ in $GL(N,\IC)$ with eigenvalues
distributed along contours in the complex plane rather than along cuts
on the real axis.
We shall leave this for future work \cite{future}.} 
with its potential given by $V(\Phi)$.
The matrix model is then defined by the partition function $Z$, which
is the following random matrix integral:
\bea
Z &=& \frac{1}{\Vol(U(N))} 
	\int [\CD\Phi]\, \exp\left( -\frac{1}{g}\,\Tr\ V(\Phi) \right),
\nn \\
&=& 
\frac{1}{\Vol(U(N))}
\int \prod_{i=1}^N d\lambda_i\,\Delta(\lambda)^2 \exp\left( -\frac{N}{g}
\sum_i V(\lambda_i) \right).
\label{Z1mat}
\eea
In eq.\ \eref{Z1mat},  we have normalized by the volume $\Vol(U(N))$ 
of the space of Hermitian matrices, and in
the second line, we have written the integration over $\Phi$
in the basis of eigenvalues.
The factor
\be
\Delta(\lambda) = \prod_{i<j} (\lambda_j - \lambda_i)
\ee
is called the Vandermonde determinant, which in the large-$N$ limit,
induces a
repulsion between the eigenvalues of $\Phi$.
The integral in eq.\ \eref{Z1mat} is performed using saddle point methods,
obtained from the variation of the integrand with respect to a single
eigenvalue $\lambda_i$.
This dictates that to $\CO(1/N)$, that is to say, in the {\em planar limit},
\be
\frac{2}{N}\sum_{j\ne i} \frac{1}{\lambda_i - \lambda_j} =
\frac{1}{g}\,V'(\lambda_i).
\label{saddle}
\ee
The value of the partition function at $\CO(1/N)$ is therefore
simply the integrand evaluated at the solutions to eq.\ \eref{saddle}, 
and hence the planar free energy is
\beq
\label{F1mat}
F := \frac{1}{N^2} \log Z
= \frac{1}{N^2}\left( 2 \log \Delta(\tilde{\lambda}_i) - 
\frac{N}{g}\sum_i V(\tilde{\lambda}_i) \right),
\eeq
with $\tilde{\lambda}_i$ the solution of eq.\ \eref{saddle}.
To solve eq.\ \eref{saddle}, we introduce the following trace, dubbed
the {\em resolvent} $R(z)$ of our matrix model:
\be
\label{defres}
R(z) := \frac{1}{N}\,\Tr\ \left[\frac{1}{z-\Phi}\right] = 
\frac{1}{N} \sum_i
\frac{1}{z - \lambda_i}.
\ee
Multiplying both sides of eq.\ \eref{saddle} by $1/(z-\lambda_i)$ and
summing over $i$, we arrive at the {\em loop equation}
\be
R(z)^2 - \frac{1}{N}\,R'(z) - \frac{1}{g}\,V'(z)R(z) 
+ \frac{1}{gN} \sum_i \frac{V'(z)-V'(\lambda_i)}{\lambda_i-z} = 0.
\label{loop}
\ee
The term involving $R'(z)$ drops out to leading order in $N$.
Additionally, in the large-$N$ limit, the {\em density of eigenvalues}
\be
\rho(\lambda) := \frac{1}{N} \sum_i \delta(\lambda-\lambda_i)
\ee
becomes continuous.
The resolvent and the normalization conditions then read
\beq
\label{res1mat}
R(z) = \int d\lambda\,\frac{\rho(\lambda)}{z-\lambda}, \qquad
\int d\lambda\,\rho(\lambda) = 1.
\eeq
In terms of the eigenvalue density, the saddle point equation \eref{saddle}
becomes
\beq
\label{sad1mat}
2 - \!\!\!\!\!\!\!\int d\tau\,\frac{\rho(\tau)}{\lambda-\tau} =
\frac{1}{g}\,V'(\lambda), 
\eeq
where by $- \!\!\!\!\!\int$ we mean principal value integration.
Likewise the loop equation \eref{loop} becomes
\beq
\label{loop2}
R(z)^2 - \frac{1}{g}\,V'(z)R(z) - \frac{1}{4g^2}\,f(z) = 0,
\eeq
with
\beq
\label{deff}
f(z) := 4g \int
d\lambda\,\frac{\rho(\lambda)}{z-\lambda}\,(V'(z)-V'(\lambda))
=
4g\left(
V'(z) R(z) - \int
d\lambda\,\frac{\rho(\lambda)V'(\lambda)}{z-\lambda}
\right).
\eeq
Eq.\ \eref{loop2} is now
an algebraic equation in the resolvent, known as the
{\em spectral curve}.

The key facts that we wish to extract from this discussion are that
knowing the functions $\rho(z)$ or $R(z)$ completely determines the
partition function $Z$, and hence the free energy $F$ and subsequently
all physical observables of the matrix model.
Finding these two functions amounts to solving eqs.\ 
\eref{res1mat} and \eref{sad1mat}.
These we recognize to be Fredholm integral equations of the first kind 
and Cauchy type, and the reader is referred to Refs.\ \cite{Ca,Kondo} for a 
solution thereof (\cf\ Ref.\ \cite{multi} in this context).
The idea is that $R(z)$ is always a multi-valued function, for which we
describe various branch cuts on the real line, the number of which is
determined by the number of critical points of $V'(z)$.
We shall shortly see that the {\em one-cut model} will be our chief
interest. 
The discontinuity across the branch cut constitutes the solutions of
eqs.\ \eref{res1mat} and \eref{sad1mat}.
In particular, the density of eigenvalues captures the multi-valued piece of 
$R(z)$,
\be
\rho(z) = \frac{1}{2\pi i} \lim_{\epsilon\to 0} \left( R(z+i\epsilon)
- R(z-i\epsilon) \right),
\label{density}
\ee
while the potential captures the holomorphic piece,
\beq
- \frac{1}{g} V'(z) = \lim_{\epsilon\to 0} \left( R(z+i\epsilon) +
R(z-i\epsilon) \right).
\label{R2V}
\eeq

\subsection{The Master Field Formulation} \label{sec:master}


At large-$N$, there is an especially convenient formulation which
computes the physical observables of the matrix model. This is the
so-called {\em Master Field} formalism \cite{Ed}.
The observables are the correlators, \ie the vacuum expectation
values (vevs), of the operators $\CO_k = \Phi^k$. These are
\beq
\left\langle \CO_k \right\rangle = Z^{-1}
\lim_{N\to\infty} \frac{1}{N} \int [\CD \Phi]\, \Tr\ \CO_k
\exp\left(-\frac{N}{g} \Tr\ V(\Phi) \right),
\eeq
with $Z$ given by eq.\ \eref{Z1mat}.
The information captured in these correlation functions may be extracted by
computing the trace of an auxiliary matrix known as the {\em master field}.
We shall quote the relevant results and refer the reader to Ref.\ \cite{gg}
for further details.

The master field offers the following remarkable properties that follow
from the application of {\em free probability theory} \cite{voi}:
\begin{itemize}
\item There is a free random variable $M$ that provides a realization of
the master field.
\item No integration needs be performed to evaluate the correlators;
it suffices to perform algebraic manipulations with
\be \label{tr}
\left\langle \CO_k \right\rangle = \tr[M^k] := \lim\limits_{N\to \infty}
\frac{1}{N} \vev{\Tr\ M^k}.
\ee
\item Any correlator in the free probability theory can be reduced to
evaluating expectation values of powers of $M$.
\end{itemize}

\newcommand{\M}{\hat{M}}

To be more explicit, the master field $M$ defines a linear functional 
$\M$ acting on a Hilbert space.
In analogy to the familiar ladder operators for harmonic oscillators, 
a basis for the Hilbert space is given by 
\be
(a^\dagger)^{n} \ket{0}.
\ee
The creation operator $a^\dagger$ is the adjoint of $a$, which annihilates
the vacuum $\ket{0}$.
The operators $a$ and $a^\dagger$ satisfy the {\em Cuntz algebra}:
\beq \label{cuntz}
a a^\dagger = \id, \quad
a^\dagger a = \ket{0}\bra{0},
\qquad
\mbox{with~~~} a\ket{0} = 0.
\eeq
This is nothing but the $q$-deformed Heisenberg algebra, 
$a a^\dagger - q\, a^\dagger a = \id$,
for the special case $q=0$.
The operator $\M$ associated to the master field $M$ is a function of 
$a$ and $a^\dagger$.
In particular, Voiculescu has shown \cite{voi} that it
is of the special form
\be \label{voic}
\M(a,a^\dagger) = a + \sum\limits_{n=0}^\infty m_n (a^\dagger)^n
\ee
with the $\{m_n\}$ some scalar coefficients of expansion.

Using eqs.\ \eref{cuntz} and \eref{voic}, the vevs may be computed by
commutations. These will turn out to be polynomials in $\{m_n\}$.
The first few such {\em \vpn}\ are
\bea 
\left\langle \CO_1 \right\rangle
	&=& \tr[M] := \vv{\M(a,a^\dagger)} = m_0, \nn \\
\left\langle \CO_2 \right\rangle
	&=& \tr[M^2] := \vv{\M(a,a^\dagger)^2} = m_0^2 + m_1, \label{poly} \\ 
\left\langle \CO_3 \right\rangle
	&=& \tr[M^3] := \vv{\M(a,a^\dagger)^3} = m_0^3 + 3 m_0 m_1 + m_2. \nn 
\eea
At each stage, a new coefficient emerges.
Thus, one can recursively compute the coefficients $\{m_n\}$ from the
traces $\{\tr[M^n]\}$ and hence all physical observables in the matrix
model.

It is convenient to map $\M(a,a^\dagger)$, which is an operator on the
Hilbert space, to a holomorphic function in the complex plane:
\be
\label{Kz}
K(z) = \frac{1}{z} + \sum\limits_{n=0}^\infty m_n z^n.
\ee
Then
\be
\label{intKz}
\vv{F'(\M(a,a^\dagger))} = \frac{1}{2\pi i} \oint_C dz\, F(K(z)),
\ee
where $F$ is any meromorphic function and
$C$ is a contour around the origin.
A proof of this statement is given in Ref.\ \cite{gg}.
Now, suppose we have found an operator $\M$ such that
\be
\tr[M^k] = \vv{\M(a,a^\dagger)^k},
\ee
for all $k$.
Then the resolvent $R(z)$, which we recall from eq.\ \eref{defres}, 
is simply the generating functional of
the moments of the matrix $M$ (\ie the vevs), becomes, due to
eq.\ \eref{intKz},
\bea
\label{RzKz1}
R(z) 
&=& \int d\lambda\, \frac{\rho(\lambda)}{z-\lambda}
= \tr\left[\frac{1}{z-M}\right]
= \sum\limits_{k=0}^\infty z^{-k-1} \tr[M^k]
= \sum\limits_{k=0}^\infty z^{-k-1} \vv{\M(a,a^\dagger)^k}   \nn \\
&=& \sum\limits_{k=0}^\infty \frac{1}{k+1} z^{-k-1} \frac{1}{2\pi i}
\oint_C d\zeta\, (K(\zeta))^{k+1} 
= -\frac{1}{2\pi i} \oint_C d\zeta\, \log[z-K(\zeta)].
\eea
Putting $K(\zeta)=\tau$, $\zeta = K^{-1}(\tau)=:H(\tau)$, we have
\be
\label{reso}
R(z) 
= -\frac{1}{2\pi i} \oint_C d\tau\, H'(\tau) \log[z-\tau] 
= \frac{1}{2\pi i} \oint_C d\tau\, \frac{H(\tau)}{\tau-z} 
= H(z) = K^{-1}(z).
\ee

The {\it point d'appui}
of the above discussion is that eq.\ \eref{reso} states that
the resolvent is the inverse, with respect to composition, of the holomorphic
function $K(z)$ obtained from taking the operator $\M(a,a^\dagger)$ and making
the formal replacements $a\mapsto 1/z$ and $(a^\dagger)^n \mapsto z^n$.
Stated briefly, {\em the resolvent is the inverse of the master field.}

Bearing this in mind, we now have an efficient method of generating the 
Voiculescu polynomials.
Take a function with a Laurent expansion as follows (which is the form
of $K(z)$),
\beq
f(z) = \frac{1}{z} + b_0 + b_1 z + b_2 z^2 + b_3 z^3 + b_4z^4 + \ldots.
\eeq
Determining the inverse of this function order by order,\footnote{
This is done as follows. To first order, let 
$f^{-1}(z) = \frac1z + \frac{x}{z^2}$, then $z = f^{-1}(f(z))$ implies
$1 + x z + b_0 z = 1 + 2 b_0 z + \CO(z^2)$, giving $x = b_0$.
Iterating to next order, we let $f^{-1}(z) = \frac1z +
\frac{b_0}{z^2} + \frac{x}{z^3}$, and matching the coefficients of
$z^2$ reads $3b_0^2 + 3 b_1 = 2(b_0^2 + b_1) + x$, giving $x = b_0^2 +
b_1$. We may continue {\it ad infinitum}.} we find that
\bea
\label{genVoi}
f^{-1}(z) &=& \frac{1}{z} + \frac{b_0}{z^2} + 
\frac{{b_0}^2 + b_1}{z^3} + \frac{{b_0}^3 + 3\,b_0\,b_1 + b_2}{z^4} + 
\frac{{b_0}^4 + 6\,{b_0}^2\,b_1 + 2\,{b_1}^2 + 4\,b_0\,b_2 + b_3}{z^5} \nn \\
&&~~~~~~~~~~~~ + \frac{{b_0}^5 + 10\,{b_0}^3\,b_1 + 10\,b_0\,{b_1}^2 + 
                 10\,{b_0}^2\,b_2 + 5\,b_1\,b_2 + 5\,b_0\,b_3 + b_4}{z^6} + \ldots.
\eea
We see the \vpn\ emerging naturally as the coefficients of expansion of
the inverse.
This is how one conveniently generates the expressions in eq.\ \eref{poly}.


Let us entice the reader with a brief preview.
We shall explore a particular one-matrix model using the technology of
the master field.
Our interest in this model is closely related to the formal substitutions
we have made in defining $K(z)$.
{\em We observe that the form of the Klein invariant \jfn\ is precisely
that of the master field of a one-matrix model, and
the coefficients of the \jfn\ determine the \vpn.}
We need only make the formal replacement $q^{-1}\mapsto a$ and $q^n\mapsto
(a^\dagger)^n$ in the Laurent expansion of the \jfn.\footnote{
Since $q^{-1}\cdot q \mapsto a a^\dagger = \id$
while $q\cdot q^{-1} \mapsto a^\dagger a = \ket{0}\bra{0}$,
this is certainly not a homomorphism.}
This paper explores features of the matrix model thus obtained.

\subsection{$\CN=1$ Field Theory, Matrix Models, and Special Geometry}
\label{sec:dvrev}

Our final ingredient comes from string theory.
Recently, Dijkgraaf and Vafa \cite{DV1,DV2,DV3,DGLVZ,CDSW} 
have made the remarkable
observation that the holomorphic data, including the superpotential, of
an $\CN=1$ theory in four dimensions can be computed from an auxiliary
matrix model.
The matrix model encodes both perturbative and non-perturbative aspects of 
the field theory.
We shall summarize a few salient features of the correspondence.
For simplicity, we examine an $U(n)$ gauge theory with an adjoint chiral
multiplet $\Phi$ and tree-level single-trace superpotential
\beq
\label{wtree}
\Wtree(\Phi) = \sum_{k=1}^{p+1} \frac{1}{k}\,g_k \Tr\ \Phi^k.
\eeq

\begin{itemize}

\item An effective low-energy 
action for $\Phi$ can be written in terms of the glueball superfield
\be
\CS = \frac{1}{32\pi^2}\ \Tr\ \CW^\alpha\CW_\alpha, 
\ee 
where $\CW_\alpha$ is the gauge field strength of the $U(n)$ vector superfield.
The theory is treated perturbatively in terms of Feynman diagrams, which
are considered using 't Hooft's double line notation.
Only the planar graphs contribute to the free energy!
The full non-perturbative superpotential in terms of the glueball is
\be
\Weff(\CS) = 
n \frac{\pa}{\pa \CS} F_0(\CS)
+ \CS (n \log (\CS/\Lambda^3)  - 2\pi i\tau ), 
\label{superp}
\ee
where $F_0(\CS)$ is the planar (genus zero) free energy
and $\tau = \frac{\theta}{2\pi} + \frac{4\pi i}{g^2}$ is the complexified
coupling.

\item The field theory results are reproduced by an associated matrix model.
We promote the chiral field $\Phi$ to an $N\times N$ Hermitian 
matrix.\footnote{Though we shall sometimes elide this distinction, the rank
of the gauge group $U(n)$ and the rank of $\Phi$ in the matrix model are
logically separate and should not be confused.
In particular,
the large-$N$ limit is {\em not} a $U(\infty)$ limit in the field theory.}
The potential of the matrix model, to be inserted into the partition function
\eref{Z1mat}, is formally the tree-level superpotential \eref{wtree}.
The planar limit in the field theory is the {\em same} as the large-$N$
limit of the matrix model.
The glueball superfield $\CS$ is identified with the 't Hooft coupling 
$gN$ in the matrix model.

\item In the original papers \cite{DV1,DV2} the correspondence between
four-dimensional $\CN=1$ field theory and matrix models is mediated through
Calabi-Yau geometry.
The theory \eref{wtree} is geometrically engineered \cite{CIV,CV} on
the (local) Calabi-Yau three-manifold
\beq
\{ u^2 + v^2 + y^2 + W'_{tree}(x)^2 = f_{p-1}(x) \} \subset \IC^4,
\label{CY}
\eeq
where $f_{p-1}(x)$, an order $(p-1)$ polynomial, is a complex deformation
parameter, to be identified with its namesake parameter $f(z)$
from eq.\ \eref{deff} that appears in the spectral curve \eref{loop2}.
Let the volume form on the Calabi-Yau be $\Omega$.
Then the following special geometry relations, as period
integrals along compact $A$-cycles and non-compact $B$-cycles,
manifestly identify the glueball $\CS$:
\bea
\CS_i = \int_{A_i} \Omega, && 
\Pi_i := \frac{\pa F_0}{\pa \CS_i} = \int_{B_i} \Omega.
\eea
The effective superpotential (\cf eq.\ \eref{superp}) \cite{GVW} is
\be
\Weff(\CS) = \int_{CY_3} G_3 \wedge \Omega = \sum_{i=1}^p N_i \Pi_i +
\alpha \sum_{i=1}^p \CS_i, \qquad
N_i := \int_{A_i} G_3, ~~~ \alpha := \int_{B_i} G_3,
\ee
where the Calabi-Yau has a three-form flux $G_3 = F_3^{(RR)} - \tau
H_3^{(NS)}$, with $\tau$ the type IIB axiodilaton.
The constant $\alpha$, independent of the choice of $B$-cycle,
is identified with the bare coupling of the low-energy Yang-Mills
theory.

Now, the non-trivial part of the Calabi-Yau threefold \eref{CY} is
the hyperelliptic curve
\be
\label{hyperellip}
y^2 = W'_{tree}(x)^2 + f_{p-1}(x),
\ee
which
encapsulates the
Seiberg-Witten geometry \cite{SW} and encodes holomorphic information about the
gauge theory.
But this curve is precisely the spectral curve \eref{loop2} 
of the corresponding matrix model!
Geometry thus interpolates between a field theory and its associated
matrix model.

\end{itemize}

\section{Modular Matrix Models} \label{sec:mmm}

Having set the stage for our analyses and the grounds for our
speculation, in this section, we will introduce our protagonist: 
the {\it modular matrix models}. We have already mentioned in
passing at the end of \sref{sec:master} the formal resemblance
between the Cuntz-expansion of the master field in a Hermitian
one-matrix model and the $q$-expansion of the \jfn, or indeed of an
abundance of modular forms. It is now our purpose to
perform the analysis in detail and construct the matrix model whose
master field is the \jfn. We shall name this model, and any 
similar models for other modular forms with analogous
$q$-expansion, the ``modular matrix model.''

\subsection{Seeking the Master}

We recall from eq.\ \eref{fourier} that the \jfn\ affords the $q$-expansion
\beq
j(q) = \frac{1}{q} + \sum_{n=0}^\infty m_n q^n
= \frac{1}{q} + m_0 + m_1 q + \ldots
\eeq
with coefficients
\bea
&&\left\{m_0, m_1, \ldots, m_6, \ldots \right\} =  \nn \\
&&~~~~~~~
\left\{744, 196884, 21493760, 864299970, 20245856256, 333202640600,
4252023300096, \ldots \right\}. \nn \\
\eea
Next, we recall that the $\IC$-number version of the master field,
defined as $K(z)$ in eq.\ \eref{Kz} is
$K(z) = \frac{1}{z} + \sum\limits_{n=0}^\infty m_n z^n$.
{\it Caveat emptur!} We are here establishing an identification not
with $K(z)$ as was done in eq.\ \eref{Kz}, 
but rather with $K(q)$, where $q = e^{2 \pi i z}$, \ie
\beq
j(q) \simeq K(q) = \frac{1}{q} + \sum_{n=0}^\infty m_n q^n.
\eeq
This is a simple change of variables. We know that $j(q)$ has the
correct expansion that resembles the master field and hence can be
used in that its coefficients yield the desired vevs. The resolvent,
according to eq.\ \eref{reso}, is the inverse, being careful now to use
$q=e^{2\pi iz}$. Therefore, we must 
determine the inverse function of $j(e^{2\pi iz})$ as a function of
$z$. We shall see below that this is a well-known function.
We summarize by stating that we seek a matrix model whose resolvent is
given by
\beq
\label{Rj}
R(z) = j^{-1}(e^{2 \pi i z}).
\eeq

\subsection{The Inverse \jfn} \label{sec:inverse}

As a brief prelude, we remind the reader that the inverse function to 
$j(q)$ can readily be determined order by order:
\beq
\label{inversej}
j^{-1}(q) = \frac{1}{q} + \frac{744}{q^2} + \frac{750420}{q^3}
+ \frac{872769632}{q^4} +  \frac{1102652742882}{q^5} +
\frac{1470561136292880}{q^6} + \ldots.
\eeq
That is,
\beq
\label{jres}
j^{-1}(q) = \sum_{k=0}^{\infty} \frac{1}{q^{k+1}} a_{k-1},
\qquad q = e^{2 \pi i z},
\eeq
where $a_k = a_k(\{m_i\}_{i=0}^k)$ is the $k$-th Voiculescu polynomial
in the coefficients of the \jfn\ as given in eq.\ \eref{inversej} 
and $a_{-1} = 1$ by definition.\footnote{
The Voiculescu polynomials and thus the coefficients in the series
expansion of the inverse 
of the \jfn\ are related to the Catalan numbers, which are the number
of two-dimensional
Dyck paths, and to the proper characters of the Monster
group. We thank J.\ McKay for pointing out Ref.\ \cite{FM}.
From the matrix model perspective, the Catalan numbers arise from the counting
of
planar graphs.
}

This expansion, to which we shall return, will of course give us no
information about the cut-structure. Indeed, the previous subsection
dictates that
it is $j^{-1}(e^{2\pi i z})$ that is required.
Now, the inverse function of Klein's absolute invariant $J(e^{2\pi i
z})$ is well-known; see for example, Ref.\ \cite{inverseJ}.
We have that
\bea
\label{res1}
J^{-1}(z) &=& \frac{i}2\,
\frac{\tilde{r}(z)-\tilde{s}(z)}{\tilde{r}(z)+\tilde{s}(z)}, \\
\tilde{r}(z) &:=& \Gamma\left(\frac{5}{12}\right)^2\
{}_2F_1\left(\frac{1}{12},\frac{1}{12};\frac{1}{2};1-z\right), \\
\tilde{s}(z) &:=& 2(\sqrt{3}-2)\ \Gamma\left(\frac{11}{12}\right)^2
\sqrt{z-1}\
{}_2F_1\left(\frac{7}{12},\frac{7}{12};\frac{3}{2};1-z\right),
\eea
where $_2F_1(a,b;c;z)$ is the standard hypergeometric function and the
range is
\beq
|J^{-1}(z)| \ge 1, \quad -\frac12 \le \Real{J^{-1}(z)} \le 0 
\qquad \mbox{for~} z \in \IC.
\eeq
Subsequently, we can determine the inverse of the \jfn\ from
eq.\ \eref{j}, which we recall to be
\beq
j(e^{2 \pi  i z}) = 1728\ J(e^{\pi  i z}),
\eeq
as
\beq
\label{res}
j^{-1}(z) = i
\left(\frac{r(z)-s(z)}{r(z)+s(z)}\right), \quad 
r(z) := \tilde{r}\left(\frac{z}{1728}\right), \quad
s(z) := \tilde{s}\left(\frac{z}{1728}\right).
\eeq
Henceforth, we send $z/1728 \to z$ and use this rescaled $z$ without
ambiguity.

\subsubsection{The Branch Cuts and Multi-Valuedness}
Let us analyze the branch cuts of eq.\ \eref{res}.
$_2F_1(a,b;c;z)$ is a single-valued function on $\IC$ with a single
branch cut $(1, \infty)$, where it is continuous from below, \ie for 
$z > 1$,
\bea
&&\lim_{\epsilon \rightarrow 0} {}_2F_1(a,b;c;z-i \epsilon)
	= {}_2F_1(a,b;c;z), \nn \\
&&\lim_{\epsilon \rightarrow 0} {}_2F_1(a,b;c;z+i \epsilon)
	=
\nn \\
&&\qquad \qquad \frac{2 \pi i e^{\pi i (a + b -c)} \Gamma(c)}
	{\Gamma(c-a)\Gamma(c-b)\Gamma(a+b-c+1)}\,
	{}_2F_1(a,b;a+b-c+1;1-z)
	+ e^{2 \pi i (a+b-c)} {}_2F_1(a,b;c;z).
\nn \\
\eea
Therefore, the branch cut we take for $_2F_1(a,b;c;1-z)$, which appears
in eq.\ \eref{res} will be $(-\infty,0)$.
Also, we take the branch cut $(1, \infty)$ for $\sqrt{z-1}$. 
In conjunction, therefore, we have that
\beq
r(z) \rightarrow
\ba{|c|c|c|}
\hline
& i \epsilon & - i \epsilon \\ \hline
z \in (-\infty,0) & -e^{\frac{\pi i}{3}} r(z) + t(z) & r(z) \\ \hline
z \in (0,1) & r(z) & r(z) \\ \hline
z \in (1, \infty) & r(z) & r(z) \\ \hline
\ea\ , \qquad
t(z) := \frac{2 i e^{-\frac{\pi i}{3}} \pi^{3/2}}{\Gamma\left(\frac23\right)} 
	{}_2F_1\left(\frac{1}{12},\frac{1}{12};\frac23;z\right),
\eeq
and that
\bea
s(z) \rightarrow
\ba{|c|c|c|}
\hline
& i \epsilon & - i \epsilon \\ \hline
z \in (-\infty,0) & -e^{\frac{\pi i}{3}} s(z) + u(z) & s(z) \\ \hline
z \in (0,1) & s(z) & s(z) \\ \hline
z \in (1, \infty) &  s(z) & -s(z) \\ \hline
\ea\ , && \nn \\
u(z) :=&& 2(\sqrt{3}-2) \sqrt{z-1}\
\frac{2i e^{-\frac{\pi i}{3}} \pi^{3/2}}{\Gamma\left(\frac23\right)} 
{}_2F_1\left(\frac{7}{12},\frac{7}{12};\frac23;z\right).
\nn \\
\eea

Recalling finally the relation \eref{Rj}, we find that
\beq
R(z + i \epsilon) \pm R(z - i \epsilon) =
\left\{
\ba{ll}
i\, \frac{\w3 (s-r) + (t-u)}{-\w3 (s+r) + (t+u)} \pm i\, \frac{r-s}{r+s}, & z \in (-\infty,0); \\
(1 \pm 1)\, i\, \frac{r-s}{r+s}, & z \in (0,1); \\
i\, \frac{r - s}{r + s} \pm i\, \frac{r + s}{r - s}, & z \in (1, \infty).
\ea
\right.
\label{Rzcombo}
\eeq

\subsection{Obtaining the Matrix Model} \label{sec:jmm}
Armed with the above mathematical machinery, we are ready to develop 
the modular matrix model associated with the \jfn. First, let us find
the eigenvalue distribution.
We recall from eq.\ \eref{density} that the density of eigenvalues is
given by the difference of the resolvent along the branch cuts.
Therefore, we find, upon simplifying eq.\ \eref{Rzcombo}, that
\beq
\label{rho-j-1}
\rho(z) = 
\left\{
\ba{ll}
\frac{1}{\pi} 
\left(
\frac{s t - r u}{(r+s)(t+u-\w3(r+s))} \right), & z \in (-\infty,0); \\ 
0, & z \in (0,1); \\
\frac{1}{\pi} \left( \frac{2 r s}{s^2 - r^2} \right), & z \in (1, \infty).
\ea
\right.
\eeq

A subtlety needs to be pointed out.
The eigenvalue density eq.\ \eref{rho-j-1} is not real, which is a property
enjoyed by Hermitian matrix models.\footnote{We thank C.\ Lazaroiu for
pointing out certain subtleties involved in using Hermitian matrix
models and emphasizing to us the need to work with complex matrices in
the context of the Dijkgraaf-Vafa correspondence
\cite{CL}.
We could of course study such a model because the techniques that we
employ in this paper    
generalize.  We shall
leave such generalization to future work \cite{future}.}
We shall apply an easy cure instead.
For simplicity, we will consider only real densities, \ie Hermitian
matrix models.
Although $\rho$ is complex from $(-\infty,0)$, it is real from $(1,\infty)$. 
Let us only consider this latter region. It is actually quite
remarkable, that there is a region such that a Hermitian matrix model
could be retrieved. This truncation, \ie a restriction to the space
of large-$N$ matrices to be integrated, is certainly a common practice,
as was performed in periodic potentials such as the Gross-Witten model
\cite{gw}. 

Another caveat is that $\rho(z)$ is not bounded, which is another
desired feature. This could be mended by
choosing a normalization scheme as follows.
We multiply $\rho$ by a constant $A$ and let $\rho$ be defined from 1 to $a$
such that
\beq
\label{norm}
A \int_1^a dz\,\rho(z) = 1.
\eeq
$A$ will depend on $a$ through eq.\ \eref{norm}.
Indeed, we can take $a$ to infinity and $A$ to zero consistently at
the end of the calculation.
We will ascertain the form of $A(a)$ when we impose finitude on
physical observables later on.
In summary then, our density function, together with a plot (recall
that we have scaled $z$ by 1728), is as follows.
\beq
\label{rho}
\rho(z) = 
\left\{
\ba{ll}
0, & z < 1; \\
\frac{A(a)}{\pi} 
\left( \frac{2 r s}{s^2 - r^2} \right), & z \in (1, a).
\ea
\right.
\qquad
\ba{c} \\ {\epsfxsize=3.75in\epsfbox{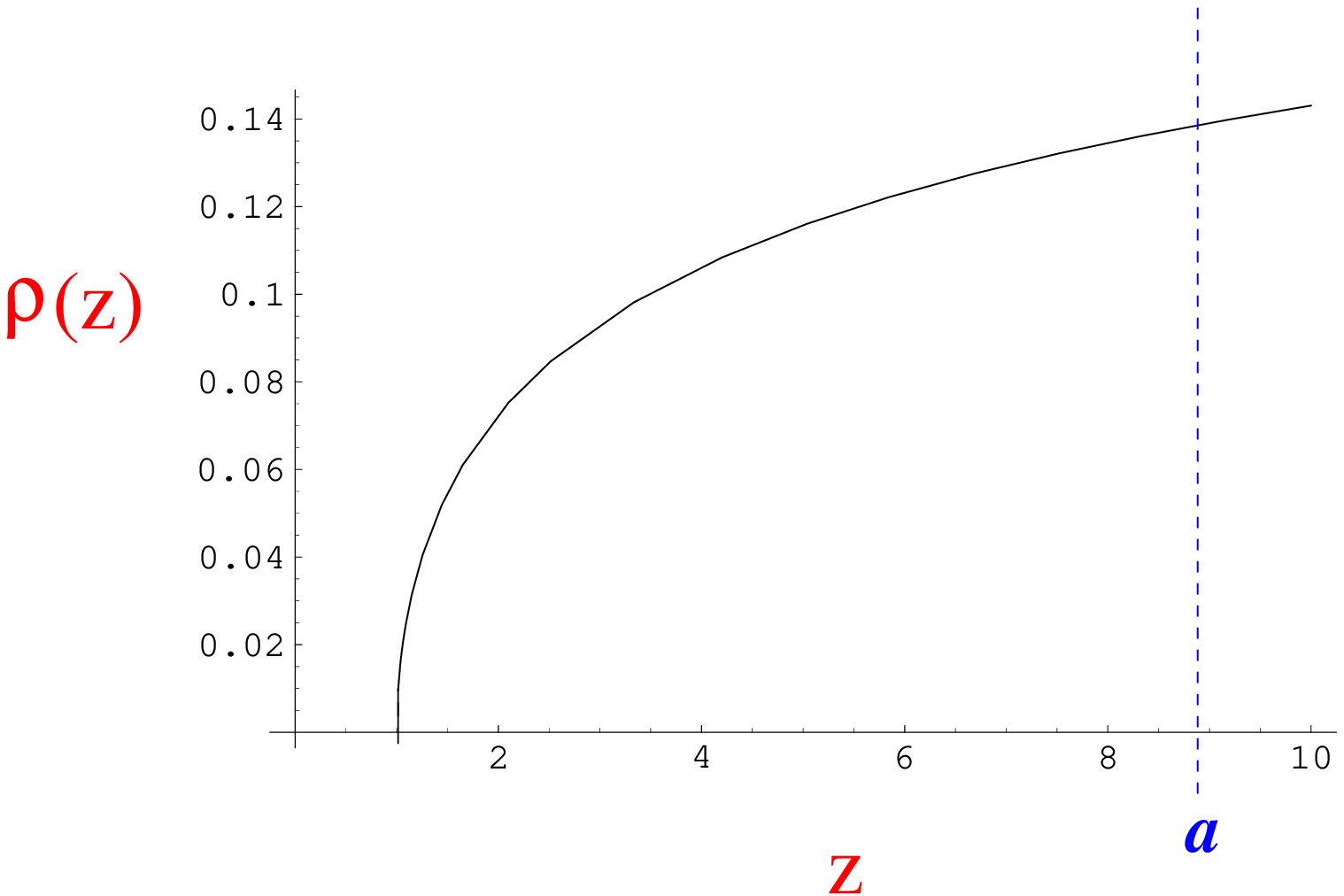}} \ea
\eeq
A benefit of this simplification is that we are now dealing only
with a one-cut matrix model instead of a two-cut model, thereby reducing
the complexity of the computations significantly.

What then is the original matrix model?
We recall from eq.\ \eref{R2V}, that, complementing the eigenvalue density,
the potential is given by the sum across the branch cuts.
Therefore, within our chosen region, we have from eq.\ \eref{Rzcombo} that
\beq
\label{Vprime}
-\frac{1}{g}V'(z) = A(a) \left(
i \frac{r - s}{r + s} + i \frac{r + s}{r - s} \right) =
A\left(j^{-1}(e^{2 \pi i z}) + \frac{1}{j^{-1}(e^{2 \pi i z})} \right), 
\quad z \in (1, a),
\eeq
which is a purely imaginary function.
Consequently,
\beq
V(z) = - g A(a)\, i \int_1^z dz'\, \left( \frac{r(z') - s(z')}{r(z') + s(z')} + 
	\frac{r(z') + s(z')}{r(z') - s(z')} \right).
\eeq
Though an analytic computation of the integral seems
intractable, $V(z)$ can be numerically determined.

To give a flavor of the form of the potential, we power expand the
first few terms as
\bea
V(z) &=&
- g A(a)\, i \left(
1.73205 + 0.335531\,z^{\frac{1}{3}} - 
  0.0649988\,z^{\frac{2}{3}} - 0.025183\,z + 
  0.0377546\,z^{\frac{4}{3}} \right. \nn \\
&&\quad 
\left. 
- 0.0151001\,z^{\frac{5}{3}} -
0.00886708\,z^2 + 0.0159885\,z^{\frac{7}{3}} - 
  0.00729369\,z^{\frac{8}{3}}+
{\cal O}(z^{3})
\right). 
\eea
The reader may be disturbed by the appearance of the fractional
powers, which simply reflect the branch point at the origin.
Now, since the minimum value of $z$ we take in our model is $z=1$, it
is more natural to expand about this point instead. Subsequently,
we arrive at an ordinary Taylor series for the potential which we
shall henceforth use:
\bea
\label{Vz}
V(z) &=&
- g A(a) i \left(2 z +
0.0696001\,\frac{(z-1)^2}{2}  - 
  0.0284334\,\frac{(z-1)^3}{3} + 
  0.0168107\,\frac{(z-1)^4}{4} - 
\right. \nn \\
&&
  0.0115703\,\frac{(z-1)^5}{5} 
+ 0.00865627\,\frac{(z-1)^6}{6} - 
  0.00682747\,\frac{(z-1)^7}{7} + \nn \\
&&\left.
  0.00558517\,\frac{(z-1)^8}{8} \right) +
{\cal O}(z^{9}). 
\eea

In possession of the eigenvalue density and the (classical) potential
in our $d=0$ matrix model, we next find the planar free energy, which
is the generating function for all the (connected) correlators.
In terms of the eigenvalue density and the potential, this
is computed in the standard way \cite{planar} by taking the large-$N$
limit of eq.\ \eref{F1mat}:
\bea
\label{freeE}
F &=& \int dx\, \rho(x) V(x) -
	\int \int dx\, dy\, \rho(x) \rho(y) \log|x-y| \nn \\
&=& \int dx\, \rho(x) \left(\frac12 V(x) - \log x\right) \nn \\
&=& \int_1^a dz\, \left(\frac{A}{\pi} \frac{2 r s}{s^2-r^2} \right)
\left(\int_1^z dz'\, (-g A\, i)\, \frac{r^2+s^2}{r^2-s^2} - \log (z)
\right).
\eea
Using the above expressions and eq.\ \eref{norm}, we can power expand
eq.\ \eref{freeE} as a function of the cut-off parameter $a$ as:
\bea
\label{freeE2}
F &=& 
- 0.108828\,i\, g\, A^2\,(a-1)^{\frac{3}{2}} - 
  \left( 0.0335904\,A + 0.0544111\,i\, g\, A^2 \right) \,
   (a-1)^{\frac{5}{2}} + \nn \\
&& \left( 0.0166887\,A + 0.00414015\,i\, g\, A^2 \right) \,
   (a-1)^{\frac{7}{2}} -
 \left( 0.00980934\,A + 0.00108499\,i\, g \,A^2 \right) \,
   (a-1)^{\frac{9}{2}} + \nn \\
&&  \left( 0.00643459\,A + 0.000424347\,i\, g \,A^2 \right) \,
   (a-1)^{\frac{11}{2}} - 
  \left( 0.00454071\,A + 0.000204799\,i\, g \,A^2 \right) \,
   (a-1)^{\frac{13}{2}} + \nn \\
&&  \left( 0.00337413\,A + 0.00011274\,i\, g \,A^2 \right) \,
   (a-1)^{\frac{15}{2}} + {\cal O}(a^8).
\eea
Now, this is an observable and needs to be finite. From 
the expansion above we can approximate the behavior of our
normalization $A$ with respect to the cut-off scale $a$.
We see that
\beq
|F| \le \sum_{n=3}^\infty |a_i| A(1+ g A) (a-1)^{n/2} \le \sqrt{a-1}\ A
	\sum_{n=1}^\infty (a-1)^{n} + {\cal O}(A^2),
\eeq
where we have used the fact that all the numerical expansion coefficients
$a_i$ in eq.\ \eref{freeE2} have modulus less than unity.
Therefore, we see that the free energy will be bounded if we choose 
the normalization
\beq
A \sim (a-1)^{-1/2} / \log(a). 
\eeq

\subsubsection{Some Salient Features of the Matrix Model} \label{sec:onmm}
 
In the above we have worked in the $z$-variable, in order to conveniently
power expand necessary physical quantities.
We must not forget however, that the interesting number theoretic
properties of the \jfn, and indeed of any modular form, are encoded in
the $q$-expansion.
If we take the potential of our matrix model, which we recall from eq.\
\eref{Vprime} is
\beq
\label{Vprime2}
V'(z) = - g A \left(j^{-1}(e^{2 \pi i z}) + \frac{1}{j^{-1}(e^{2 \pi i
z})} \right), 
\eeq
and expand about $q = \exp(2 \pi i z)$, we would find, according to
the order by order treatment of eq.\ \eref{inversej}, that
\bea
\frac{d}{dz}V(q) &=& - g A \left(
\left(
\frac{1}{q} + \frac{744}{q^2} + \frac{750420}{q^3}
+ \frac{872769632}{q^4} +  \frac{1102652742882}{q^5} +
\frac{1470561136292880}{q^6}+ \ldots
\right) \right. \nn \\
&&
\left.
+ \left(
q - 744 - \frac{196884}{q}- \frac{167975456}{q^2} -
\frac{180592706130}{q^3} - \frac{217940004309744}{q^4}+ \ldots
\right)
\right)
\nn \\
&=&
- g A \left(
q - 744 - \frac{196883}{q} -\frac{167974712}{q^2}
- \frac{180591955710}{q^3} - \frac{217939131540112}{q^4} - \ldots
\right).
\nn \\
\eea
Using $dz = \frac{1}{2\pi i}\frac{dq}{q}$, a $q$-expansion for the potential
of our Hermitian matrix model may subsequently be developed:
\beq
\label{Vq}
V(q) = \frac{- g A}{2\pi i} \left(
q - 744 \log q + \frac{196883}{q} + \frac{83987356}{q^2}+
\frac{60197568710}{q^3} + \frac{54485001077436}{q^4} + \ldots
\right).
\eeq

From eq.\ \eref{poly}, we recall that the expectation values of the
matrix model are \vpn\ of the expansion coefficients of the master field.
The master field we have chosen in our theory is the \jfn; therefore,
the expectation values computed from eq.\ \eref{Vz} or eq.\ \eref{Vq}
will be Voiculescu polynomials in the coefficients of the \jfn.
Indeed, from the coefficients of the series expansion of $j^{-1}(z)$
obtained order by order from $j(z)$ in eq.\ \eref{inversej}, we see that
\bea
\label{vevs}
744 &=& 744 = m_0; \nn \\
750420 &=& 196884 + 744^2 = m_1 + m_0^2; \nn \\
872769632 &=& 21493760+3\cdot 744\cdot 196884 + 744^3 = 
	m_2 + 3 m_0 m_1 + m_0^3; \nn \\
20245856256 &=& 864299970 + 2\cdot 196884^2 + 4\cdot 744 \cdot
	21493760 + 6\cdot 744^2 \cdot 196884 + 744^4 \nn \\
&&	= m_3 + 2m_1^2 + 4m_0m_2 + 6m_0^2m_1 + m_0^4; \nn \\
\ldots
\eea
which is the expected pattern of Voiculescu polynomials.
If we compute the fat Feynman diagrams associated with the
correlation functions
\beq
\tr[\Phi^p] = \vev{0|\hat\Phi^p|0} = 
Z^{-1} \int [\CD\Phi]\, \Tr\ \Phi^p \exp\left(-\frac{1}{g} \Tr\ V(\Phi) \right),
\eeq
using $V(\Phi)$ from eq.\ \eref{Vz} or eq.\ \eref{Vq}, we shall retrieve
precisely the list in eq.\ \eref{vevs}.

\subsubsection{A Closely Related Matrix Model} \label{sec:othermm}

Equally could we have asked ourselves, now that we have a Hermitian
one-matrix model whose vacuum expectation values are Voiculescu polynomials
in the $q$-coefficients of the \jfn, how might we establish a similar model
whose moments are the $q$-coefficients themselves?
To construct this latter model, we need only invert the Voiculescu
polynomials.
Therefore, we desire a master field
\beq
\label{Mz}
M(q(z)) = \frac{1}{q} + \sum_{n=0}^\infty \mu_n q^n,
\eeq
such that
\bea
\tr[\Phi] &=& \mu_0 = 744; \nn \\
\tr[\Phi^2] &=& \mu_1 + \mu_0^2 = 196884; \nn \\
\tr[\Phi^3] &=& \mu_2 + 3 \mu_0 \mu_1 + \mu_0^3 = 21493760; \nn \\
\tr[\Phi^4] &=& \mu_3 + 2\mu_1^2 + 4 \mu_0 \mu_2 + 6 \mu_0^2 \mu_1 + \mu_0^4 
	=  864299970; \nn \\
\tr[\Phi^5] &=& \mu_4 + 5 \mu_1 \mu_2 + 5 \mu_0 \mu_3 + 10 \mu_0 \mu_1^2 +
	10 \mu_0^2\mu_2 + 10 \mu_0^3\mu_1 + \mu_0^5 =
	20245856256; \nn \\
\ldots.
\eea
Subsequently, the first few coefficients can be iteratively obtained:
\beq
\mu_0 = 744, ~~~ \mu_1 = -356652, ~~~ \mu_2 = 405710240, ~~~ \mu_3 = -582814446942, 
~~~ \ldots. 
\label{mus}
\eeq
In other words, we impose that
\beq
j(q) = \frac{1}{q} + \mu_0 + (\mu_1 + \mu_0^2)q + (\mu_2 + 3 \mu_0
\mu_1 + \mu_0^3)q^2 + \ldots,
\eeq
signifying that
\beq
\frac{1}{q^2}\, j(1/q) = \frac{1}{q} + \frac{\mu_0}{q^2} + 
	\frac{\mu_1 + \mu_0^2}{q^3} + \frac{\mu_2 + 3 \mu_0
	\mu_1 + \mu_0^3}{q^4} + \ldots.
\eeq
Therefore, using eq.\ \eref{genVoi}, this means that
\beq
\frac{1}{q^2}\, j(1/q) = M(q)^{-1},
\eeq
and hence we have constructed the inverse for our master field
eq.\ \eref{Mz}, which, we recall, is the resolvent of our desired matrix
model.
In terms of $z$, we have that
\beq
M(e^{2 \pi i z})^{-1} = e^{-4 \pi i z} j(e^{- 2 \pi i z}).
\eeq
Thus,
\beq
\label{Rztrial}
R(z) = M(z)^{-1} = \frac{1}{2 \pi i} \log \left(\frac{1}{z^2}\, j(1/z) \right) 
= -\frac{\log z}{\pi i} + \frac{1}{2\pi i}\log j(1/z).
\eeq
An analogous route to the preceding sections could be taken.
Now, as $j(z)$ has no branch points, the only branch cut comes
from the logarithm, which extends from 0 to $\infty$. 
However, $j(z)$ is not defined over the real axis, but only on
the upper half plane.
Therefore, this deceptively simpler model escapes the usual analysis in the
context of Hermitian matrix models.
Our initial choice, wherein the vacuum expectation values are the Voiculescu
polynomials in the $q$-coefficients on which we have thus far focused,
is a more becoming choice.
Of course, we could employ the technology of more involved (\eg complex)
matrix models on eq.\ \eref{Rztrial}, but that is another story, which
we shall relate another time.
This is a lesson in the simple fact that not every meromorphic function
can be used as a resolvent of a Hermitian matrix model; that the inverse
of the \jfn\ could be, and is consistently constructable as one, is a
pleasant surprise.

\subsubsection{Variations on a Theme by the Master} \label{sec:vars}

The content of Moonshine subsists in the observation that the coefficients
in the $q$-expansion \eref{fourier} encode data about the irreducible
representations of the Monster sporadic group.
{\em Any} resolvent that organizes this information faithfully will offer
a realization of the Monster group's character table in terms of a matrix
model and can then be employed to make statements in the language of field
theory and geometry about the Monster sporadic group itself.
While the relations \eref{Rj} and \eref{Rztrial} are particularly natural
in the sense that they express the resolvent or the \vpn\ in terms of the
\jfn\ in a convenient and direct manner, there are other models that also
bottle the substance of Moonshine.
For example, 
we could have taken $R(z) = \exp(2 \pi i j^{-1}(z))$ as our starting point.
That we have made certain choices in our analysis is not to imply that
these are the only ones that exist.
However, the path we have chosen, to consider $R(z) = j^{-1}(e^{2\pi i z})$,
is particularly organic. In this case,
we are led to a zero-dimensional, Hermitian one-matrix model with a single
branch cut that lends itself to simple
computations and still distills some essential
virtues of Moonshine.

\subsection{A Dijkgraaf-Vafa Perspective} \label{sec:dv}

\subsubsection{An ${\CN} = 1$ Gauge Theory} \label{sec:susy}

Having constructed a Hermitian matrix model which naturally encodes
the \jfn, it will be expedient to seek yet more physical quantities
which bear 
connexion with $j(z)$. We shall present some intriguing speculations
in this section.

As reviewed in \sref{sec:dvrev},
it is the realization of a recent set of seminal works by Dijkgraaf and Vafa
\cite{DV1,DV2,DV3} that matrix models of the type described above actually
compute the full non-perturbative content of an ${\cal N} = 1$ gauge
theory. 
The tree-level superpotential, 
according to the correspondence, 
is simply the (classical) potential for the matrix
model. 
We formally replace the Hermitian matrix by the single adjoint field
$\Phi$ of the $U(n)$ ${\cal N} = 1$ gauge theory, and obtain the
superpotential as
\bea
\label{Vtree}
W_{tree}(\Phi) &=&
- i\, g\, A \left(2 \Phi +
0.0696001\,\frac{(\Phi-1)^2}{2}  - 
  0.0284334\,\frac{(\Phi-1)^3}{3} + 
  0.0168107\,\frac{(\Phi-1)^4}{4} - 
\right. \nn \\
&&
\left.
  0.0115703\,\frac{(\Phi-1)^5}{5} 
+ 0.00865627\,\frac{(\Phi-1)^6}{6} - 
  0.00682747\,\frac{(\Phi-1)^7}{7} + \ldots \right) \nn \\
&=&
\frac{- g A}{2\pi i} \left(
e^{2 \pi i \Phi} - 744\,(2 \pi i \Phi) + 196883\,e^{-2 \pi i \Phi} + 
83987356\,e^{-4 \pi i \Phi} + \right. \nn \\
&&
\left.
~~~~~~~~~~~60197568710\,e^{-6 \pi i \Phi} + 54485001077436\,e^{-8 \pi i
\Phi} + \ldots 
\right).
\eea
using eqs.\ \eref{Vz} and \eref{Vq}. The fact that the tree-level potential
is non-polynomial need not disturb us; the famous Gross-Witten model \cite{gw}
has a cosine potential, for which a Dijkgraaf-Vafa analysis was
carried out in Ref.\ \cite{DV2}.

We can use the standard Dijkgraaf-Vafa prescription to determine the full
non-perturbative potential in terms of the glueball condensate
\beq
\CS = \frac{1}{32\pi^2} \Tr\ \CW_\alpha \CW^\alpha,
\eeq
which obtains from
the gauge field strength $\CW_\alpha$.
Accordingly, one identifies $\CS$ with the 't Hooft coupling in the matrix
model, and the full superpotential is simply
\beq
\Weff \simeq \CS \log \CS + \diff{~}{\CS} F_0
\eeq
using the planar free energy $F_0$ computed in eq.\ \eref{freeE2}.
It is now necessary to restore the powers of $N$ in the matrix model,
accompanying the powers of the gauge coupling $g$, on which the
cut-parameter $a$ depends through normalization. The derivative 
becomes, from eq.\ \eref{freeE2},
\bea
\label{diffF0}
&&\diff{~}{\CS} F_0 = 
- 0.163242\, i\, g\, A'(\CS)^2 a'(\CS) (a(\CS) - 1)^{\frac12} + \nn \\
&&\left(0.014491\, i\, g \,{A(\CS)}^2\,a'(\CS) + 
  A(\CS)\,\left( 0.058411\,a'(\CS) + 
     0.10882\,i \,A'(\CS) \right)  - 0.03359\,A'(\CS) \right) (a(\CS) -
1)^{\frac52} + \nn \\
&&\left(
0.0023339\,i\, g \,{A(\CS)}^2\,a'(\CS) + 
  A(\CS)\,\left( 0.03539\,a'(\CS) + 
     0.00217\, i \,A'(\CS) \right)  - 0.00981\,A'(\CS)
\right)(a(\CS) - 1)^{\frac92} + \nn \\
&& + \CO((a(\CS)-1)^{\frac{13}{2}}),
\eea
where $a'(\CS)$ and $A'(\CS)$ are derivatives of $a$ and $A$ with respect
to $\CS$. The functional form of $a(S)$ is easily determined by
restoring the coupling and further, the appropriate power of $N$,
into eq.\ \eref{norm}, \viz
\beq
\frac{A}{gN} \int_1^a dz\,\rho(z) = 1.
\eeq
This means that, upon taking the derivative of $\CS = gN$,
\beq
\diff{~}{\CS} f(a) = \frac{\pi}{A} 
\left( 1 - \frac{\CS}{A} \frac{\pa A}{\pa\CS} \right),
\eeq
where $f(a)$ is the anti-derivative of $\frac{2 r(z) s(z)}{s(z)^2 -
r(z)^2}$. 
We can then numerically solve this differential equation
to find the form of $a(\CS)$ and back-substitute into
eq.\ \eref{diffF0} to determine the effective superpotential in terms of the
glueball $\CS$. 
The $\Weff$ we obtain at last, has a $q$-expansion just like the potential
$V$ did. A pressing question is whether this is a modular form. 
Unfortunately, 
due to the fact that we cannot integrate eq.\ \eref{Vprime}
analytically, the answer to this question eludes us at present.
As we only have the perturbative
expansion of the function, it is difficult to determine such
properties as modularity. Be that as it may, were $\Weff$ to be a
modular form, 
it would provide a natural quantum generalization of the \jfn.

\subsubsection{A Calabi-Yau Geometry} \label{sec:cy}

Our final venture, having embarked upon a course from the
$q$-expansion of the \jfn, shall be to the realm of special geometry.
We recall from \sref{sec:dvrev} that the Dijkgraaf-Vafa correspondence
also provides us a singular Calabi-Yau geometry associated to the matrix
model which is a generalization of a conifold.
The singular geometry is that of hyperelliptic curve, given by the spectral
curve of the eigenvalue densities, transverse to a $\IC^2$.
In particular, embedded in $\IC^4$, the Calabi-Yau is
\beq
\{u^2 + v^2 + s(x,y) = 0 \} \subset \IC[x,y,u,v],
\eeq
with $s(x,y)$ the hyperelliptic spectral curve.
For our model, this curve is
\beq
\label{spec}
y^2 - V'(x)^2 - \frac{1}{4g^2}f(x) = 0,
\eeq
with $f(x)$ the remainder term given in the spectral curve \eref{loop2}.
It is to be considered as a quantum correction (deformation) to the
geometry.

Now our potential, from eq.\ \eref{Vprime}, is
$V'(z) = A\left( j^{-1}(e^{2 \pi i z}) + \frac{1}{j^{-1}(e^{2 \pi i
z})} \right)$. Therefore, 
eq.\ \eref{deff} tells us that we have
\bea
f(z) &=&
4g\left(
V'(z) R(z) - 
\int_1^\infty d\lambda\, \frac{\rho(\lambda)V'(\lambda)}{z-\lambda}
\right) \nn \\
&=&
- 4Ag^2\left(
1 + (j^{-1}(e^{2 \pi i z}))^2 + i
\int_1^\infty
\frac{d\lambda}{z-\lambda}
\left[
\frac{A}{\pi} \frac{4 r s(r^2+s^2)}{(s^2 - r^2)^2}
\right]
\right).
\eea
Hence, the spectral curve is fully determined, complete with its
quantum correction $f(z)$.

\newcommand{\ch}{{\rm ch}}

The above geometrical digression is reminiscent of an earlier
proposal circulated amongst mathematicians.
It is a known result (\cf Ref.\ \cite{Doran}) that for the family of
elliptic curves over $\IP^1$ given by
\beq
\label{unif}
y^2 = 4x^3 - \frac{27s}{s-1}x-\frac{27s}{s-1}, \quad s \in \IP^1,
\eeq
the mirror map $z : \IP^1 \mapsto s \in \IP^1$ at a point of maximal
nilpotent monodromy, is precisely
\beq
z(q) = \frac{1}{j(q)},
\eeq
the reciprocal of the \jfn, namely the inverse of $j(q)$ with respect to 
multiplication.
In fact, an old conjecture of Lian and Yau \cite{LY}
states that for a wide class of K3 surfaces, the $q$-series of the
mirror map is a Thompson series $T_g(q)$ for some element $g$ of the
Monster sporadic group $\Mon$.\footnote{A {\em Hauptmodul} is a function
$J_g(q) = q^{-1} + \sum_{n=1}^\infty a_n(g) q^n$, for $g\in\Mon$.
Although $|\Mon| \simeq 8\times 10^{53}$ there are ``only'' 171 distinct 
Hauptmoduls of $\Mon$.
Consider $J_\id(q) = j(q) - 744$.
The coefficients of the \jfn\ (and also $J_\id(q)$) are related to the 
irreducible representations of $\Mon$.
Replacing the $m$-th coefficient of the Fourier expansion of the Hauptmodul
$J_\id(q)$ by the corresponding representation of $\Mon$, we have a formal
power series
$$ 
T_\id(q) = V_{-1} q^{-1} + 0 + V_1 q + V_2 q^2 + \ldots,  
$$
where $V_r$ are {\em head representations} of $\Mon$ 
($V_{-1} = {\bf 1}$, $V_1 = {\bf 1} \oplus {\bf 196883}$, etc.) 
and $V = \bigoplus_r V_r$.
Thompson \cite{th} proposed that by analogy, for an arbitrary element 
$g\in\Mon$, we write the formal power series
$$ 
T_g(q) := \ch_{V,q}(g) = \ch_{V_{-1}}(g) q^{-1} + 0 + \ch_{V_1}(g) q +
\ch_{V_2}(g) q^2 + \ldots,
$$
where $\ch_{V_r}(g)$ is a group character of $\Mon$.}
What is the relation between our hyperelliptic curve \eref{spec} and
the family \eref{unif} of elliptic curves?
We leave this as an open problem.

\section{Prospectus} \label{sec:spec}

Grounded in the observation that 
the formal resemblance between the $q$-expansion of Klein's elliptic \jfn\
and the Cuntz-expansion of the master field in a one-matrix model is more
than an accident, we have in this paper constructed the 
``modular matrix model.''
The key property of this Hermitian bosonic matrix model is that its
master field is by construction $j(q)$.
The vacuum expectation values, \ie the observables, in the theory are
Voiculescu polynomials in the $q$-coefficients of the \jfn.
We have computed the eigenvalue density, the resolvent, and the free
energy of the model.
Furthermore, we have established the (classical) potential.
Utilizing the Dijkgraaf-Vafa correspondence, we have also constructed
the associated supersymmetric $\CN=1$ gauge theory which localizes to
such a bosonic matrix model.
Subsequently, we have computed the full non-perturbative superpotential
as well as a local Calabi-Yau geometry as a fibration of a hyperelliptic
curve on which the gauge theory may be geometrically engineered.

A host of tantalizing questions immediately opens to us.
We have much alluded to the complex (\ie $GL(N,\BC)$) version of our analysis,
which, among others, would give us a matrix model directly encoding the
$q$-coefficients of $j(q)$.
Does such an extension lead to further intriguing observations?
Of course, one cannot resist considering other variations of the matrix
model. The current resurgence of interest in $c=1$ matrix models perhaps
tempts us to seek the one-dimensional extension to the matrix models that we
have considered here. Would this quantum mechanics of matrices, by promoting
$\Phi$ to $\Phi(t)$ in eq.\ \eref{Vtree}, offer new perspectives on the \jfn?

It is clear that
we should explore the special geometry of the hyperelliptic
curve for the modular matrix model further.
Can one establish a relation between our hyperelliptic spectral curve and
the Lian-Yau conjecture concerning the family of elliptic K3 surfaces?
Now, the latter conjecture relates classes of K3 surfaces to the Thompson
series of Hauptmoduls.
What about our Calabi-Yau three-fold, which describes a deformed conifold
geometry?
How does our geometry, which naturally encodes the \jfn, relate to these
Thompson series?

Perhaps the most tempting speculation is what we shall call ``Quantum
Moonshine.''
Our analyses have given us a matrix model corresponding to the Klein
invariant \jfn.
Now because of Moonshine, our modular matrix model, in encoding the
\jfn, further encodes the dimensions of the irreducible representations
of the Monster group.
The manifestation of Moonshine, in its relation to vertex operators,
has already appeared in string theory.
In ``Beauty and the Beast'' \cite{DGH}, bosonic
closed string theory compactified on an orbifold of the Leech lattice with
a background Neveu-Schwarz $B$-field was shown to have a partition
function that encodes the Monster representations. How does our story
enter into this picture?

Already in eqs.\ \eref{Vq} and \eref{mus} we have encountered some 
quite large integers that are suggestive of this interplay.
It was
suggested to us\footnote{We thank A.~Iqbal for pointing this out to
us.}
that the Voiculescu polynomials in the $q$-coefficients may count
certain open-string topological invariants associated with the
large-$N$ geometry. Moreover,
as we have emphasized, we have performed all our computations in the planar
limit, which is to say, we have solved the saddle point equation to $\CO(1/N)$.
The large-$N$ limit of this matrix model allows us, by the Dijkgraaf-Vafa
correspondence, to recover details about an $\CN=1$ field theory whose
tree-level superpotential is determined by the resolvent.

But this is only a leading order result.
One can compute the higher $\CO(1/N^k)$ corrections in the matrix model.
In the field theory, this corresponds to non-planar diagrams and
gravitational corrections.
Such higher-order terms arise through quantum effects and should proffer
a quantum modification to the Conway-Norton Moonshine conjectures.
In other words, the $\CO(1/N^k)$ terms, in correcting the observables
in the matrix model, and hence the Calabi-Yau geometry, should provide a
natural generalization of the elliptic $j$-invariant and hence provide a
quantum version of the Monster group.
What interpretation do the quantum corrections have in terms of the Monster
sporadic group?
Is this a quantum deformation to the presentation of the Monster?
We have mentioned in \sref{sec:dv} that the full non-perturbative
superpotential might be a modular form, which would be yet another way
of finding quantum corrections to Moonshine.
The Dijkgraaf-Vafa program therefore would offer not only a perturbative
window into non-perturbative physics, but also a perturbative window into
number theory.

Finally, we see that as the form of $q$-expansion is generic to modular
forms, what would an analogous analysis to ours give for one of a myriad 
other modular functions?
What number theoretic and geometric data would these encode?
We have perhaps raised as many questions as we have presented answers,
and this is our hope.
It is our desire that ``modular matrix models'' should shed light
into various hitherto unexplored corners of string theory, number theory,
and geometry.

\vspace{1in}
\section*{Acknowledgments}
{\it F\"ur Gustav Mahler.\\}

We are grateful for enlightening discussions with Ian Ellwood, Amer
Iqbal,
Calin Lazaroiu, John McKay,
Djordje Minic, and Mithat Unsal.
We are especially indebted to Asad Naqvi, for his keen physical insights,
and to Matthew Szczesny, for our ongoing collaborations.
The authors thank the High Energy Group at the University of Pennsylvania
and the Physics Department at Virginia Tech for sponsoring reciprocal
visits to each other's home institutions.
VJ thanks the Pacific Institute for Mathematical Sciences at the University
of British Columbia for providing the stimulating environment in which this
paper was completed.
Finally, we toast to stately, plump Buck Buchanan for
offering us a strong drink 
in celebration of Bloomsday.

The research of YHH is funded in part by cooperative research agreement
\#DE-FG02-95ER40893 with the U.~S.~Department of Energy and the National
Science Foundation Focused Research Grant DMS0139799 for ``The Geometry
of Superstrings''.  The research of VJ is funded in part by
cooperative
research agreement \#DE-FG05-92ER40709 with the U.~S.~Department of Energy.

\end{document}